%Paper: hep-th/9309108
%From: li@het.brown.edu (Miao Li)
%Date: Mon, 20 Sep 93 15:01:21 EDT
%Date (revised): Wed, 22 Sep 93 15:36:12 EDT

%This paper uses harvmac

\input harvmac

\Title{BROWN-HET-923, hep-th/9309108}{Dirichlet String Theory and Singular
Random Surfaces}
\centerline{Miao Li\foot{E-mail: li@het.brown.edu}}
\bigskip
\centerline{Department of Physics}
\centerline{Brown University}
\centerline{Providence, RI 02912}
\bigskip
We show that string theory with Dirichlet boundaries is equivalent to
string theory containing surfaces with certain singular points.
Surface curvature is singular at these points. A singular point is resolved
in conformal coordinates to a circle with Dirichlet boundary conditions.
We also show that moduli parameters of singular surfaces coincide with those
of smooth surfaces with boundaries. Singular surfaces with saddle points
indeed arise in the strong coupling expansion in lattice QCD. The kind
of saddle point, which may be the origin of a singular point we need, is of
infinite order.

\Date{9/93}
%\draft

Green proposed to study string theory propagating with surfaces bordered
by Dirichlet boundaries, in order to reproduce the high energy behavior
of scattering amplitudes in QCD \ref\green{M.B. Green, Phys. Lett. B201
(1988) 42; B266 (1991) 325; B282 (1992) 380.}. Earlier references in the
subject include those in \ref\early{J.H. Schwarz, Nucl. Phys. B65 (1973) 131;
E. Corrigan and D. Fairlie, Nucl. Phys.
B91 (1975) 527; M. Green, Nucl. Phys. B103 (1976) 333.}.
For a recent approach to this
string theory, see \ref\mli{M. Li, Dirichlet strings, BROWN-HET-915,
hep-th/9307122.}. Is there any evidence based upon more theoretical grounds
in support of the Dirichlet string theory, other than merely requiring
account for ``phenomenological'' high energy results? The purpose of this
note is to provide such evidence.

Starting with QCD, either from 't Hooft's $1/N$ expansion
\ref\thooft{G. 't Hooft, Nucl. Phys. B72
(1974) 461.} or from Wilson's strong coupling expansion on the lattice
\ref\wilson{K. Wilson, Phys. Rev. D8 (1974) 2445.}, one arrives at the
conclusion that QCD may be described by a non-interacting string theory
at the large $N$ limit. Unfortunately this turned out not quite to be the case
\ref\wein{D. Weingarten, Phys. Lett. B90 (1980) 285.}, at least some
modifications are needed. Kazakov proposed a modified strong coupling
expansion in \ref\kazakov{V. Kazakov, Phys. Lett. B128 (1983) 316; JETP 85
(1983) 1887.}, in which not only more degrees of freedom are required, but also
some singular saddle points appear on random surfaces. This approach was
further developed in \ref\kostov{I. Kostov, Phys. Lett. B138 (1984) 191; B147
(1984)445; K.H. O'Brien and J.-B. Zuber, Nucl. Phys. B253 (1985) 621;
Phys. Lett. B144 (1984) 407.}, for a recent review, see \ref\kaz{V. Kazakov,
Lectures given in the Trieste Spring School and Workshop-1993 on String
Theory, hep-th/9308135.}. Despite subtleties in going from the strong coupling
regime to the weak coupling regime, those saddle points nevertheless survive.
It is these singular points which interest us here. We shall argue that these
points could serve as the origin of Dirichlet boundaries.

In the strong coupling expansion, a saddle point arises when $2n$ ($n>1$)
plaquettes are contracted cyclically on a link. The deficit angle at this
point is $2\pi(n-1)$.
If this point survives the continuum limit, this deficit angle will induce a
singular curvature, i.e., the curvature blows up approaching this point.
Associated to a saddle point of a given order, there is a coefficient
depending on the type of the saddle point.
For the simplest saddle point with $n=2$, there is an additional weight
factor $1/N$. There are higher order corrections in terms of $1/N$. These
can be explained as the coalescing of several saddle points of the same
order. Now let us assume that this singularity
can be described by a singular intrinsic curvature on the continuum surface.
For a finite $n$, the curvature behaves as a $\delta$-function centering at
the saddle point. This kind of singularity will turn out to be too mild to
induce a Dirichlet boundary. Nevertheless let us discuss this singularity
first. Note that a saddle point of finite order is precisely a branch point
of the same order found in the QCD$_2$ string theory \ref\gross{D.J. Gross and
W. Taylor, Nucl. Phys. B400 (1993) 181; B403 (1993) 395; J. Minahan, Phys. Rev.
D47 (1993) 3430.}. To see this, consider a metric $ds^2=|z|^{2(n-1)}
|dz|^2$. The curvature of this metric behaves as $\sqrt{g}R\sim \delta^2(z)$.
Use coordinates $w=z^n$ instead, the metric becomes smooth $ds^2=|dw|^2$.
The singular property is encoded in the fact that the $z$ plane covers the
$w$ plane $n$ times, and $z=0$ is a branch point of order $n$. It is
instructive to define the deficit angle by
\eqn\deficit{
{\int_C ds\over \int_0^r ds}-2\pi,}
where $C$ is the contour with a constant radius $r$. The first term is just
the ratio of the circumference to the radius. In polar coordinates $z=
re^{i\phi}$,
the metric reads $ds^2=r^{2(n-1)}(r^2d\phi^2+dr^2)$. It is easy to check that
our definition of the deficit angle yields a value $2\pi(n-1)$ in this case.
To study an appropriate conformal field theory, which is needed in a string
theory, on a closed surface with such a branch point, one may either resolve
this mild singularity by going to a higher genus (as done in the QCD$_2$
string theory), or one may simply insert an appropriate vertex operator
\ref\knizhnik{see for example, V.G. Knizhnik, Commun. Math. Phys. 112 (1987)
567.} at this point and deal with a smooth surface (in coordinates $w$).

Now the kind of singular point inducing a Dirichlet boundary is different
from a saddle point of any finite order. The reason is that the curvature
blows up badly at the singular point. Without loss of generality, let us
consider a metric
%\foot{This metric is considered in \ref\dpoly{D. Polyakov,
%Nucl. Phys. B401 (1993) 698.}, where $a$ is assumed to be an integer.}
%
\eqn\sing{ds^2=r^{2a}d\phi^2+dr^2.}
This metric is singular at $r=0$ when $a<1$. We also require $a>0$,
otherwise $r=0$ is
not really a point. A simple calculation yields $R=2a(1-a)r^{-2}$. This
curvature is so singular that the integral $\int\sqrt{g}R$ diverges at $r=0$.
Note that we have to put the restriction $a<1$, otherwise $\int\sqrt{g}R$
converges. Thus, this singular point can not be resolved by going to higher
genus. This is not surprising, since the deficit angle as defined in
\deficit\ is $2\pi (r^{a-1}-1)$, which approaches infinity as $r$ approaches
zero. If we are to explain this singularity as originating from saddle
points, it must consist of infinitely many saddle points on the same link with
arbitrary large order.
We conjecture that this is the case, although it is hard to have a derivation
directly from the strong coupling expansion. After all the strong coupling
expansion has not been proven to have a continuum limit yet.

We shall not assume any particular conformal field theory (CFT) in the QCD
string theory here. To study a CFT with geometry \sing, it is necessary
to find a coordinate system in which the metric reads $\hbox{exp}(\phi)
|dz|^2$. Then we need to study this CFT in the geometry $|dz|^2$ by a Weyl
rescaling. It is straightforward to see that the metric \sing\ can be
written as $r^{2a}|dw|^2$, where $w={1\over 1-a}r^{1-a}+i\phi$. Now $w$
furnishes complex coordinates on the semi-infinite cylinder, and the boundary
of this cylinder is at $r=0$. Notice that if $a=1$, one can not use these
coordinates. The semi-infinite cylinder can be further mapped to the complex
plane with a unit disk removed, by $z=\hbox{exp}(w)$. The metric reads in
these coordinates
$$ds^2={1\over |z|^2}\left({1-a\over 2}\hbox{ln}|z|^2\right)^{{2a\over 1-a}}
|dz|^2.$$
The Weyl factor $\hbox{exp}(\phi)$ vanishes on the boundary $|z|=1$. The
Liouville action, resulting from a Weyl rescaling, is infinite with this
Weyl factor. We have seen that in a conformal coordinate system, the singular
point {\it necessarily} blows up to a circle. If the CFT to be studied
includes spacetime coordinates $X^\mu$ as a subset of conformal fields, the
boundary conditions for these fields are Dirichlet, as this boundary represents
just a previous singular point of curvature. We also note that the resulting
conformal geometry is independent of parameter $a$. Formally, the Liouville
action
depends on $a$, however, we shall assume that this Liouville action be
eventually canceled out by that from contribution of $b$-$c$ ghosts. We
postpone discussion of boundary conditions for these ghosts to a little later.

The above consideration of resolving a singular point can be generalized to
the sphere with a singular point, and in general to a sphere with many
such singular points. An interesting question then arises. If any such singular
geometry is resolved resulting in a sphere with boundaries, are there enough
configurations of singular geometries to cover the moduli space of the
conformal sphere with boundaries? The answer to this question is positive.
We consider the sphere with two singular points as an example. Assume that
$r=0$ and $r=\infty$ be two singular points on the sphere, in polar
coordinates. The metric is approximately as in \sing\ around $r=0$, with
parameter $a_1$. The metric is approximately $r'^{2a_2}d\phi^2+dr'^2$
around $r=1/r'=\infty$. $a_i$ are all positive and less than one. We can
write down such a metric readily
\eqn\twos{ds^2=(1+r^2)^{-2}\left((r^{2a_1}+r^{4-2a_2})d\phi^2+dr^2\right),}
$r^{2a_1}$ dominates $r^{4-2a_2}$ when $r\rightarrow 0$; while $r^{4-2a_2}$
dominates $r^{2a_1}$ when $r\rightarrow \infty$. The above metric is written
in conformal coordinates as
\eqn\conf{ds^2=(1+r^2)^{-2}(r^{2a_1}+r^{4-2a_2}){|dz|^2\over |z|^2},}
where
$$z=e^{R+i\phi},\quad R(r)=\int_0^r {dr\over (r^{2a_1}+r^{4-2a_2})^{1/2}}.$$
Apparently $R$ is a well-defined function of $r$. The singular point $r=0$ is
mapped to $|z|=1$, the unit circle. The singular point $r=\infty$ is mapped
to the circle $|z|=\hbox{exp}(R_0)>1$. $R_0$ is finite, and given
by
\eqn\modu{R_0=\int^\infty_0{dr\over (r^{2a_1}+r^{4-2a_2})^{1/2}}=
{1\over \sqrt{\pi}(4-2a_1-2a_2)}\Gamma({1-a_1\over 4-2a_1-2a_2})\Gamma
({1-a_2\over 4-2a_1-2a_2}).}
We thus see that the sphere with two singular points is mapped to a conformal
sphere with two disks removed, and it is just the annulus. The single
real moduli parameter is the radius of the outer boundary $\hbox{exp}(R_0)$,
which is finite. The moduli parameter depends on both $a_1$ and $a_2$.
Adjusting $a_i$, $R_0$ can be sent to infinity. However,
$R_0$ can never be zero, so geometries as described by \twos\ can never
cover the whole moduli space. To cover the whole moduli space, one can
replace the factor $r^{2a_1}+r^{4-2a_2}$ in \twos\ by $(r^{2a_1}+r^{4-2a_2})
\rho^2(r)$. Here $\rho(r)$ is a function with $\rho(0)=\rho(\infty)=1$. Now
$R_0$ is not determined by \modu\ but by
$$R_0=\int_0^\infty {dr\over (r^{2a_1}+r^{4-2a_2})^{1/2}\rho(r).}$$
It is possible to adjust $\rho(r)$ to let $R_0$ approach zero. This is done
by increasing $\rho(r)$ for all $r$ except $r=0, \infty$. Geometrically,
the result of doing
this is that the circumference of a circle with constant $r$ is increased.
Notice that the distance between two points $r=0$ and $r=\infty$ remains the
same. This picture then explains why in conformal geometry, the outer
boundary becomes closer and closer to the inner boundary.  The whole
moduli space is covered in this way. The moduli
parameter then depends not only on parameters $a_i$, but also on details
of geometry. However, the existence of boundaries is due to the existence of
singular points.

These considerations generalize to a sphere with many singular points. It is
hard and unnecessary to work out details of mapping singular geometries
to conformal geometries with boundaries. We shall only argue that the whole
moduli space of the latter is covered by varying singular geometries. Suppose
there are $N\ge 3$ singular points, therefore there are $N$ parameters
$a_i$ characterizing these singularities. Together with the positions of these
singularities, there are $3N$ real parameters available. Since there is a
global conformal invariance group $SL(2,C)$ on the sphere, only $2N-6$
parameters
out of $2N$ position parameters are relevant. So we are left with $3N-6$ real
parameters. This number is precisely the dimension of the moduli space
of the sphere with $N$ components of boundary. As in the $N=2$ case,
moduli parameters are complicated functions of the parameters of singular
geometries, and it may be impossible to cover the whole moduli space if we
consider only one class of geometries, such as that in \twos. The moduli
space can be covered by considering all possible singular geometries, again
as in the $N=2$ case. We believe that the conclusion presented in this section
applies equally to Riemann surfaces of higher genus with singular points.

After having shown that Dirichlet string theory is equivalent to a closed
string theory with singular surfaces, we come to the issue of $b$-$c$ ghosts.
If the boundary conditions for scalar fields $X^\mu$ are Dirichlet, what
are the boundary conditions for $b$-$c$? Apparently, we can not impose
Dirichlet boundary conditions on $b$-$c$ consistently. These fields are
fermionic, so Dirichlet boundary conditions can either be $b=0$ or
$c=0$ on the boundary. If $b=0$, then there is no nontrivial solution to
the equation of motion $\overline{\partial}b=0$, this would contradict the
existence of moduli parameters. If $c=0$, the trivial solution to $\overline
{\partial}c=0$ contradicts the existence of  nontrivial conformal
Killing vectors on, say, the disk. So we are left only with the possibility
of Neumann type boundary conditions, namely $c\partial=\bar{c}\overline{
\partial}$ and $b(dz)^2=\bar{b}(d\bar{z})^2$ on the boundary.
We now show how this is consistent with consideration of singular points.
Consider the ghost fields in the geometry \sing. For the $c$ field, there
are two components $(c^r,c^\phi)$ in polar coordinates. In conformal
coordinates $(z,\bar{z})$, the two components become $c^z=zr^{-a}c^r+izc^\phi$
and $c^{\bar{z}}=\bar{z}r^{-a}c^r-i\bar{z}c^\phi$. So both $c^z$ and $c^{
\bar{z}}$ become singular on the boundary, if $c^r\ne 0$. If we demand that
$c^r$ approach zero faster than $r^a$ (which approaches zero more
slowly than $r$), we find the desirable Neumann boundary conditions. As for
the $b$ field, we find that two terms $b_{rr}(\partial r/\partial z)^2$
and $b_{r\phi}(\partial r/\partial z)(\partial \phi/\partial z)$ in $b_{zz}$
are vanishing
when $r\rightarrow 0$, for $\partial r/\partial z$ approaches zero.
Neumann boundary conditions for this field then follow.

We showed in \mli\ that the Liouville action resulting from a Dirichlet
boundary is essentially the same as the one from a Neumann boundary.
In other words, Dirichlet boundary conditions do not change the central
charge of the CFT. With a $b$-$c$ ghost system obeying Neumann boundary
conditions,
it is possible to cancel the Liouville action from the matter CFT part.
Together with our discussion above on the singular surfaces, we conclude that
study of a string theory with such surfaces is made possible by studying the
Dirichlet string theory. Because the latter makes perfect sense, it is
then a reasonable hypothesis that singular surfaces contribute significantly
in certain circumstances, such as in the QCD string theory.

As argued by
Green, as well as in \mli\ for a general case, the Dirichlet boundaries
provide us with a mechanism to generate the appropriate high energy behavior
observed in strong interactions, which an ordinary string theory fails to do.
Why is this so from the standpoint of the strong coupling expansion?
The rest of this note is devoted to a plausible answer to this question.

First, let us remind ourselves how scattering amplitudes at high energies
can be calculated in the pure glue theory. Color singlet states are
supposed to be glueballs. A glueball state, as an eigen-state of four
momentum, is a superposition of many states with different numbers of gluons.
Determination of the wave function of a particular glueball state is a
nonperturbative problem. Fortunately, as a high energy process at a wide angle
is concerned, there must be a factorization theorem, similar to the one for
meson scattering \ref\fact{see articles in Perturbative Quantum
Chromodynamics, ed. A.H. Mueller, World Scientific, 1989.}. This theorem
states that the scattering amplitude
factorizes into two factors. One of them has to do with the structure of the
glueball, which may be called the distribution amplitude;
another is the scattering amplitude between two individual gluons. The
latter can be calculated perturbatively. Even the dependence of the
distribution amplitude on the momentum transfer can be calculated
perturbatively. The dependence of the whole amplitude on the large momentum
transfer is largely determined by the two gluon scattering  amplitude.
This is translated into the short distance behavior of the two point
function of the gauge potential $\langle A_\mu(x_1)A_\nu(x_2)\rangle$, which
is calculated perturbatively.

Instead of considering high energy scattering amplitudes of glueballs, one
can consider correlation functions of gauge invariant local operators
constructed from field strength.
High energy behavior is translated into certain short distance behavior of
these correlation functions. As far as any short distance behavior is
concerned, one can use the operator product expansion, again
calculated perturbatively \ref\asymp{H.D. Politzer, Phys. Rep. 14C (1974)
131.}

In the modified strong coupling expansion of Kazakov \kazakov, a Lagrangian
multiplier $\alpha$, a hermitian matrix, is introduced to enforce unitarity
of the link variable. Weight of each saddle point is the expectation value
of a negative momentum of $\alpha$, namely $W_n=\langle {1\over N}\hbox{tr}
\alpha^{-2n}\rangle$, where $n$ is the order of the saddle point. It can be
shown in the one plaquette model that solutions of the weight exist in both
the strong coupling phase and the weak coupling phase \kaz. We are certainly
interested in the weak coupling phase. Here one finds that for large $\beta$,
the inverse of the coupling constant, $W_n$ goes to $\beta^{-2n}$. This factor
cancels the factor $\beta^{2n}$ from area of $2n$ plaquettes contracted on
the link. This shows that all saddle points of higher order contribute
in the weak coupling phase.

Similar arguments lead to demonstration of the existence of the weak coupling
phase in QCD$_4$ \kaz, the physical model we are mostly interested in.
Curiously, Kazakov showed that the weight $W_n$ is associated to local
condensation of the gluon field. Absorbing a factor $\beta^{2n}$ into the
weight, he finds
\eqn\cond{W_n=\beta\langle {1\over N}\hbox{tr}F_{\mu\nu}^2\rangle +\dots}
The quantity on the r.h.s. of \cond\ can be calculated perturbatively in
the weak coupling regime, using the standard Feynman rules. Now consider
correlation functions of gauge invariant local operators which in turn are
constructed from plaquette variables, in the framework of strong coupling
expansion. Saddle points appear in any such correlation function, and the
weight in the weak coupling phase is given by eq.\cond. When two operators
in the correlation approach each other, we expect that saddle points
near these operators play a more and more important role. It is then expected
that the operator product expansion in the lattice context is dominated by
saddle points, this is consistent with the fact that the weight in
eq.\cond\ is determined perturbatively.

A lesson about the underlying CFT in the QCD string theory is readily drawn
from the above discussion. In addition to the power-law behavior in the
coefficients of the operator product expansion, there are logarithmic
corrections, as functions of $\hbox{ln}(p^2/\Lambda^2_{QCD})$. The QCD scale
$\Lambda_{QCD}$ is the only free parameter in QCD. This parameter is
replaced by the string tension $T\sim (\alpha')^{-1}$ in the string theory.
So it is expected that similar factors involving $\hbox{ln}(p^2/T)$ appear
in the string calculations, even when only one singular point on the surface
is present. With the choice of the simplest CFT described by free fields
$X^\mu$ on the world sheet, it is shown in \green\ and \mli\ that no
such logarithmic factors are obtained. Thus, some modified CFT is required
to yield logarithmic corrections. This is in contrast to the belief that
logarithmic corrections arise from renormalizing infinities in the string
theory. If this were true, one would be forced to introduce another independent
scale. It is questionable that such a scale can be related to the string
tension in any reasonable manner. From the strong coupling expansion, it is
obvious that the free CFT is not the right theory, for example back-tracking
configurations are not suppressed in this theory. It is likely that the correct
CFT for the QCD string, if any, must involve interacting fields $X^\mu$ and
additional fermionic fields to give logarithmic corrections as well as
to suppress back-tracking.

In conclusion, no matter what the CFT may look like in the QCD string theory,
singular surfaces must be taken into account in order to produce power-law
behavior
in high energy scattering amplitudes. We have shown that singular points
can be resolved by introducing Dirichlet boundaries, so the theory can be
studied systematically.

\noindent{\bf Acknowledgments}

We would like to thank C.-I Tan for spiritual support, and for asking
the question if it is possible to understand the heterotic pomeron in some
string theory. We would also like to thank B. Urosevic, W. Zhao for
comments, and A.T. Sornborger for reading the manuscript. This work was
supported by DOE grant DE-FG02-91ER40688-Task A.

\listrefs\end